\documentclass{emulateapj}
\usepackage{natbib}
\newcommand{\kms}{\,{\rm km\,s^{-1}}}
\newcommand{\msun}{\,{\rm M_\odot}}

\newcommand{\beq}{\begin{equation}}
\newcommand{\eeq}{\end{equation}}
\newcommand{\ba}{\begin{eqnarray}}
\newcommand{\ea}{\end{eqnarray}}
\def\spose#1{\hbox to 0pt{#1\hss}}
\newcommand{\lta}{\mathrel{\spose{\lower 3pt\hbox{$\mathchar"218$}}
      \raise 2.0pt\hbox{$\mathchar"13C$}}}
\newcommand{\gta}{\mathrel{\spose{\lower 3pt\hbox{$\mathchar"218$}}
      \raise 2.0pt\hbox{$\mathchar"13E$}}}
\def\simlt{\mathrel{\rlap{\lower 3pt\hbox{$\sim$}}\raise 2.0pt\hbox{$<$}}}
\def\simgt{\mathrel{\rlap{\lower 3pt\hbox{$\sim$}} \raise 2.0pt\hbox{$>$}}}
\lefthead{Volonteri et al. }
\righthead{Accretion onto quiescent black holes}

\begin{document}
\title{Massive black holes in stellar systems: `quiescent' accretion and luminosity}
\author{M. Volonteri\altaffilmark{1}, M. Dotti \altaffilmark{2}, D. Campbell\altaffilmark{1} \& M. Mateo\altaffilmark{1},}
\altaffiltext{1}{University of Michigan, Astronomy Department, Ann Arbor, MI, 48109}
\altaffiltext{2}{Max Planck Institute for Astrophysics, Karl-Schwarzschild-Str. 1,  85741 Garching, Germany}

\begin{abstract}
Only a small fraction of local galaxies harbor an accreting black hole, classified as an active galactic nucleus (AGN). However, many stellar systems are plausibly expected to host black holes, from globular clusters to nuclear star clusters, to massive galaxies. The mere presence of stars in the vicinity of a black hole provides a source of fuel via mass loss of evolved stars. In this paper we assess the expected luminosities of black holes embedded in stellar systems of different sizes and properties, spanning a large range of masses. We model the distribution of stars and derive the amount of gas available to a central black hole through a geometrical model. We estimate the luminosity of the black holes under simple, but physically grounded, assumptions on the accretion flow. Finally we discuss the detectability of `quiescent' black holes in the local Universe.

\end{abstract}


\section{Introduction}
Dynamical evidence indicates that massive black holes  with masses in the range $M_{BH} \sim 10^6-10^9\,\msun$ ordinarily dwell in the centers of most nearby galaxies \citep{ferrarese2005}. The evidence is particularly compelling in the case of our own galaxy, hosting a central black hole with mass $\simeq 4\times10^6\,M_\odot$   \citep[e.g.,][]{Schodel2003,Ghez2005}.   Massive black holes with smaller masses exist as well.  For example, the Seyfert  galaxies, POX  52 and NGC 4395, are thought to contain massive black holes with mass $\sim 10^5\,M_\odot$ \citep{Barth2004,Peterson2005}.  
Low mass black holes might also exist in dwarf galaxies, for instance in Milky Way satellites.  If these black holes exist they can help us understand the process that formed the seeds of the massive holes we detect in much larger galaxies \citep{svanwas2010}. Black holes in massive galaxies have a high probability that the central black hole is not ``pristine", that is, it has increased its mass by accretion or mergers. Dwarf galaxies undergo instead a quieter merger history, and as a result, if they host black holes, they still retain some ``memory'' of the original seed mass distribution \citep{VLN2008}.

The dynamical-mass estimates indicate that, across a wide range, central black hole mass are about 0.1\% of the spheroidal component of the host galaxy, with a possible mild dependence on mass  \citep{Magorrian1998,MarconiHunt2003,Haring2004}.  A tight correlation is also observed between the massive black hole mass and the stellar velocity dispersion of the hot stellar  component \citep[``M-$\sigma$",][]{Ferrarese2000,Gebhardt2000, Tremaine2002, Graham2008,Gultekin2009}. \cite {Lauer2007} suggest that at least some of these correlations break down at the largest galaxy and black hole masses \citep[but see][]{Bernardi2007,Tundo2007,Graham2008}. One unanswered question is whether this symbiosis extends down to the lowest galaxy and black hole masses \citep{Greene2008}, due to changes in the accretion properties \citep{Mathur2005}, dynamical effects \citep{Volonteri2007}, or a cosmic bias \citep{VN09,svanwas2010}.  

It has also been proposed \citep[e.g.,][]{PZ2004,Ato2004} that black holes of intermediate mass (between the stellar mass range, $\sim$ few tens $\msun$, and the supermassive black hole range, $\gta 10^5\msun$), can form in the center of dense young star clusters. It is proposed that the formation of the black hole is fostered by the tendency of the most massive stars to concentrate into the cluster core through mass segregation.  The merging of main-sequence stars via direct physical collisions can enter into a runaway phase, forming a very massive star, which can then collapse to form a black hole  \citep{Begelman1978, ebisuzaki2001,Miller2002,PZ2002,PZ2004,freitag2006a,freitag2006b,Ato2004,ato2006}.  Observational evidences for intermediate mass black holes in globular clusters are scant \citep[e.g.,][and references therein]{vandermarel2010,Pasquato2010}. Dynamical measurements are hampered by the small size of the sphere of influence of these black holes, and only four candidates have currently been identified, in M15, M54, G1 and $\omega$ Centauri \citep{Gerssen2002,Ibata2009,Gebhardt2005,Noyola2008}. The radio and X-ray emission detected from G1 make this cluster the strongest candidate, although alternative explanations, such as an X-ray binary are possible \citep{Ulvestad2007,Pooley2006}. 

`Massive' black holes (more massive than stellar mass black holes) are therefore expected to be widespread in stellar systems, from those of the lowest to highest mass. Only a small fraction of these massive black holes are active at levels that are expected for AGNs, and, indeed, most massive black holes at the present day are `quiescent'. However, because MBHs are embedded in stellar systems, they are unlikely to ever become completely inactive. A massive black hole surrounded by stars could be accreting material, either stripped from a companion star or available as recycled material via mass loss of evolved stars.  \citep{Ciotti1997}. \cite{Quataert2004} model the gas supply in the central parsec of the Galactic center due to the latter process. Winds from massive stars can provide $\sim 10^{-3}\msun {\rm yr^{-1}}$ of gas, with a few percent, $\sim 10^{-5}\msun {\rm yr^{-1}}$, of the gas flowing in toward the central massive black hole. \cite{Quataert2004} shows that the observed luminosity from Sgr A* can indeed be explained by relatively inefficient accretion of gas originating from stellar winds.

Elliptical galaxies with quiescent massive black holes, systems for which we have both accurate massive black hole masses and data about their surroundings, hint that stellar winds may be a significant source of fuel for the massive black hole.  The hot gas of the interstellar medium, lending itself to X-ray observations, cannot be the sole source of fuel for at least some massive black holes. In particular, some massive black holes are brighter than one would expect for inefficient accretion, but significantly less bright than for normal accretion \citep{Soria2006a}.  The X-ray luminosity can vary by \(\sim3\) orders of magnitude displaying no relationship between massive black hole mass or the Bondi accretion rate \citep{pellegrini2005}.  It is likely that warm gas that has not yet been thermalized or virialized originating from stellar winds and supernovae from near the massive black hole provides a significant amount of material for accretion, possibly an order of magnitude larger than the Bondi accretion rate of hot interstellar medium gas alone \citep{Soria2006b}.

We attempt in this paper a simple estimate of how much recycled gas is available for accretion onto a massive black hole in different stellar systems, from globular clusters to galaxies, including dwarf spheroidals, nuclear star clusters in the cores of late type galaxies and early type normal galaxies. We show that the amount of fuel available to massive black holes through stellar winds in quiescent galaxies is indeed meager, and unless extreme conditions are met, X-ray detection of massive black holes in globular clusters and low-mass galaxies is expected to be uncommon. 

\section{Method}

\subsection{Stellar models}
To model the accretion rate, we must  choose  3-dimensional stellar distributions for the various stellar systems we consider here. 
For globular clusters and dwarf  spheroidals we assume the stars to be distributed following a
Plummer profile: 
\beq
\rho(r)=\frac{3}{4\pi}\frac{M_{\rm
    stellar}}{a^3}\left(1+\frac{r^2}{a^2}\right)^{-5/2}, 
\label{eq:plummer}
\eeq 
where $a=R_{\rm eff}$ is the core radius. 

Early type galaxies and nuclear clusters are modeled as Hernquist spheres:
\beq
\rho(r)=\frac{M_{\rm
    stellar}}{2\pi}\frac{r_h}{r(r+r_h)^3},
\label{eq:hernquist}
\eeq 
where the scale length $r_h\approx R_{\rm eff}/1.81$.  To fully define the stellar systems we have only to relate the stellar mass, $M_{\rm stellar}$, to the effective radius, $R_{\rm eff}$.


 For globular clusters, we recall that simulations by \cite{Baumgardt2005,Baumgardt2004} suggest that globular clusters with massive black holes have relatively large cores $a\sim 1-3$ pc \citep[see also][]{Trenti2007}. Consistent results were found using Monte Carlo simulations \citep{Umbreit2009} and in analytical models \citep {Heggie2007}.  The core radii (where measured) of globular clusters hosting intermediate mass black hole candidates, are roughly consistent with the values we considered, ranging from approx 0.5 pc in M15 (Gerssen et al. 2002, core radius from the catalog presented in Harris
et al. 2010\footnote{http://physwww.physics.mcmaster.ca/~harris/mwgc.dat}), up to few pc in omega Centauri (Noyola et al. 2008).

For early type galaxies, we adopt the fits by \cite{Shen2003} for stellar-mass vs effective radius in Sloan Digital Sky Survey galaxies:
\begin{equation}
      	R_{\rm eff}=2.5\,\left(\frac{M_{\rm stellar}}{4\times10^{10}\msun}\right)^{0.56}\, {\rm kpc}.
\end{equation}      
The scatter is roughly 0.2 dex for stellar
masses between $10^8\msun$ and $10^{10}\msun$: $\sigma_{\ln
  R}=0.34+0.13/[1+(M_{\rm stellar}/4\times10^{10}\msun)]$.

We note that for 5 galaxies (NGC~4697, NGC~3377, NGC~4564, NGC~5845, NGC~821) where measurements of the effective radius are
available (along with stellar masses, black hole masses, and gas density- see \cite{Soria2006a} and \cite{MarconiHunt2003}) the fits derived by \cite{Shen2003} provide values of the effective radius roughly 55\% times larger than the measured value. This is likely due to  \cite{Shen2003} definition of effective radius as the radius enclosing 50 per cent of the Petrosian flux.  This definition differs from the standard definition of projected radius enclosing half of the total luminosity. We therefore scale the fit for early type galaxies by a factor of 0.55 for consistency.  As shown below (Fig.~\ref{size}) this small correction does not influence the accretion rate we derive. 

For dwarf spheroidals, we fit the data presented in \cite{Walker2009, Walker2010}.  We assume a constant mass-to-light 
ratio of two for the visible component, and derive stellar masses from the total luminosities: 
\beq
R_{\rm eff}=0.93\,\left(\frac{M_{\rm stellar}}{10^7
  \msun}\right)^{0.36}\, {\rm kpc}, 
\eeq 
where the uncertainties in
the slope and in the normalization are 0.06 and 0.2 dex respectively.
Finally for nuclear clusters we fit the stellar mass vs effective
radius data presented in \cite{Seth2008}, leading to: 
\beq R_{\rm eff}=7.9 \times 10^{-3}\,\left(\frac{M_{\rm stellar}}{10^7
  \msun}\right)^{0.17}\, {\rm kpc}, 
\eeq 
where the uncertainties in
the slope and in the normalization are 0.05 and 0.3 dex respectively.
These scalings are shown in Figure~1.

\begin{figure}
\plotone{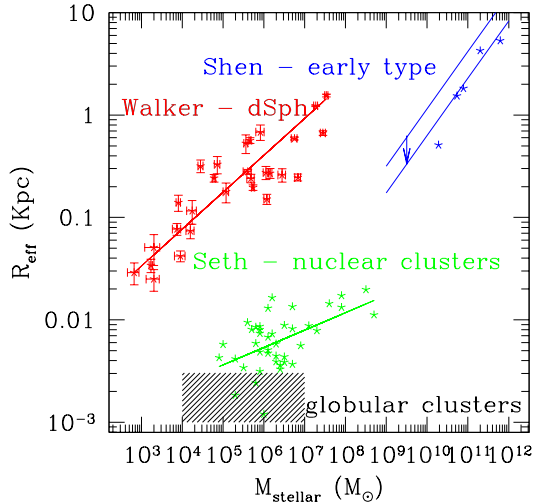}
\caption{\footnotesize Relationship between half-mass radii and
  stellar mass for different galaxy morphological types. For dwarf
  spheroidals and nuclear clusters we show the data along with our best fit. 
  For elliptical galaxies we show the effective radii of 5 galaxies from  \cite{Soria2006a}, 
  along with Shen et al. (2003) fit and a correction of a factor 0.55. 
  We include as a shaded area the range in half-mass radii and
  stellar mass adopted for globular clusters.}
\bigskip
\label{radii}
\end{figure}

%

\subsection{Geometrical model}

We develop here a simple geometrical model to estimate the accretion
rate onto a massive black hole in a stellar system, fueled by mass loss from stars
\citep{Quataert1999}. If a star is located at a
distance $r$ from the massive black hole, and if it produces an isotropic wind, with
velocity $v_{\rm wind}$, only the fraction of gas which passes within
the accretion radius of the massive black hole, 
\beq R_{\rm acc}=2GM_{\rm BH}/(v^2_{\rm wind}+\sigma^2+c^2_s), \eeq 
  
  can be accreted (ignoring
gravitational focusing).  Here $\sigma^2=GM_{\rm stellar}/(2.66r_h)$ is the velocity dispersion of
the stellar system at the half-mass radius.  For a Hernquist profile, where the density in the inner region $\rho \propto r^{-1}$,
the velocity dispersion decreases towards the center. Estimating $\sigma$ at the half mass radius gives a conservative lower 
limit to the accretion radius, and hence the accretion rate. 
Following \cite{Miller2002},  we assume that in equation 6 the sound speed $c_s= 10 \kms$, and, $v_{\rm wind} = 50 \kms$
as reference values, although we study the effect that a different $v_{\rm wind}$ has on our model (see Figure~\ref{vw}).

If $\sigma \gg v_{\rm wind}$, $R_{\rm acc}$ depends only on the properties of the potential well of
the stellar distribution, not on the
wind properties. In particular, $R_{\rm acc} \simeq M_{\rm BH}R_{\rm
  eff}/M_{\rm stellar} \simeq 10^{-3}R_{\rm eff}$ if $M_{\rm
  stellar}=10^3 M_{BH}$. 
  Note that, at fixed black hole mass, the more massive the galaxy, the smaller $R_{\rm acc}$ is, as the scaling of $R_{\rm eff}$ with $M_{\rm stellar}$ is 
 a power law with exponent less than one (see, e.g., equation~3). 
    On the other hand, if $\sigma \ll v_{\rm
  wind}$, $R_{\rm acc}$ depends only on the wind velocity. These two
limits are apparent in Figures~\ref{vw} and~\ref{size}, and they will be
discussed in the next section.

Geometrical considerations suggest that, for $r>R_{\rm
  acc}$: 
\beq 
\dot M_{{\rm acc},*}=\frac{1}{2}\dot M_*\left[1-\left(1-\frac{R^2_{\rm
      acc}}{r^2}\right)^{1/2}\right], 
\label{eq:geom}
\eeq where $\dot M_*$ is the mass loss rate from the star. If the star
lies within $R_{\rm acc}$, we consider $\dot M_{{\rm acc},*}=\dot
M_*$.   Eq.~(\ref{eq:geom}) implicitly assumes that the stars have
  a spherically simmetric distribution and that their velocity field
  (and, as a consequence, the velocity field of the wind) is
  isotropic. In a rotating stellar system, the presence of net angular
  momentum of the gas can change the accretion rate onto the black
  hole \citep[e.g.,][]{Cuadra2008}. A study of the dependence of the
  accretion rate on the degree of rotational support of the stellar
  distribution is beyond the scope of this paper.

The total contribution from all stars is found by integrating over the
density profile of the stellar system:
\beq
\dot M_{\rm acc}=\int_{R_{\rm acc}}^{\infty} {4\pi
  r^2\frac{\rho(r)}{\langle m_*\rangle}\dot M_{{\rm
      acc},*} dr},
\label{eq:mdot}
\eeq 
where $\langle m_*\rangle$ is the mean stellar mass and $\rho$ is
given by eq.~(\ref{eq:plummer})~and~(\ref{eq:hernquist}).  The
normalization in eq.~(\ref{eq:mdot}) is given by the cumulative mass
loss rate of all the stars in the stellar structure, that we estimate
following \cite{Ciotti1991}: 
\beq 
\dot M_{\rm gal}=1.5\times 10^{-11}\msun {\rm
  yr^{-1}}\frac{L_B}{L_{B,\odot}} \left(\frac{t_*}{15 {\rm Gyr}}\right)^{-1.3}, 
\eeq 
where $t_*$ is the age of the stellar population, and $L_B$ is the total luminosity of the 
stellar system. We set $t_*=5$ Gyr for dSphs and nuclear star clusters, and $t_*=12$ Gyr for early type galaxies and globular clusters. 
We derive B-band luminosities from stellar masses assuming a mass-to-light ratio of 5 in the B-band.

We obtain an upper limit of the luminosity of the massive black hole by assuming that
the whole $\dot M_{\rm acc}$ is indeed accreted by the massive black hole.


\subsection{Accretion rate and luminosity}

\begin{figure}
\plotone{{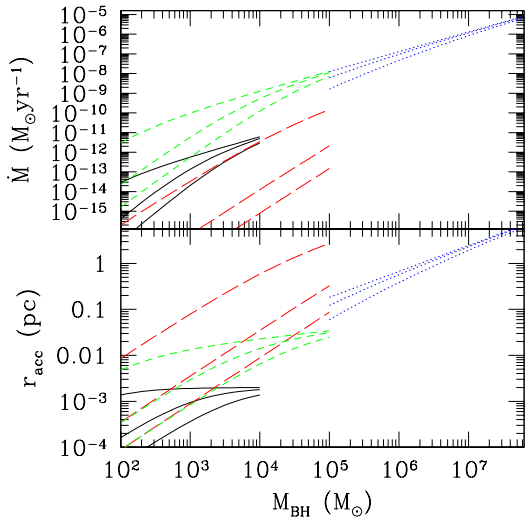}}
\caption{\footnotesize Top panel: accretion rate, in solar masses per year, onto a black hole in a stellar system with $M_{\rm stellar}=10^3 M_{BH}$. In each set of curves the wind velocity varies from 100$\kms$ (bottom) to 50 $\kms$ (middle) to 10 $\kms$ (top). Solid curves: globular clusters. Long-dashed curves: dwarf spheroidals. Short-dashed curves: nuclear star clusters. Dotted curves: early-type galaxies. Bottom panel: accretion radius for the same systems.  
}
\label{vw}
\end{figure}

Figure ~\ref{vw} shows the resulting accretion rate for a central massive black hole
in different stellar systems, where we assume that the massive black hole mass scales
with the mass of stellar component, $M_{\rm BH}=10^{-3} M_{\rm
  stellar}$ \citep{MarconiHunt2003,Haring2004}, and we have considered
$v_{\rm wind}$ a free parameter. We have assumed that $R_{\rm
  eff}$ scales exactly with $M_{\rm stellar}$ following the
relationships discussed above.  Note that for high values of the
  stellar masses in early-type galaxies and nuclear star clusters, the
  accretion rate and $R_{\rm acc}$ do not depend on the wind
  velocities. In these cases $\sigma\gg v_{\rm wind}$, and the
  accretion rate depends only on the properties of the host stellar
  structure and on the black hole mass (see the discussion of Equation~6 above). 

\begin{figure}
\plotone{{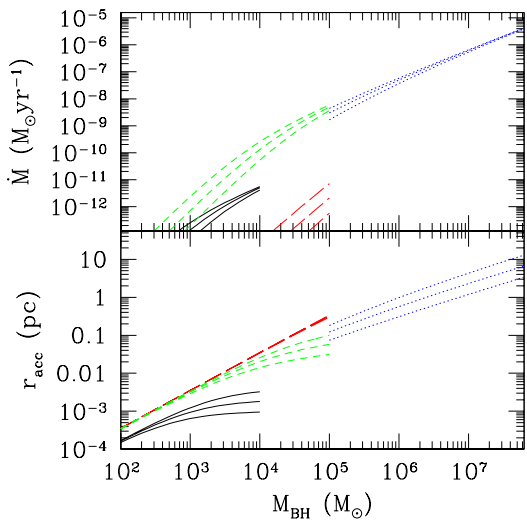}}
\caption{\footnotesize Top panel: accretion rate, in solar masses per year, onto a black hole in a stellar system with $M_{\rm stellar}=10^3 M_{BH}$. In each set of curves we vary the size of the stellar system. For
globular clusters we assume $R_{\rm eff}=1$ pc (top); $R_{\rm eff}=2$ pc (middle); $R_{\rm eff}=4$
pc (bottom).
For galaxies, the middle curve shows the best fit $R_{\rm eff}$ at a given stellar mass (Equations 4, 5 and 6), the top curves assume that $R_{\rm eff}$ is half the best fit value, the bottom curves that $R_{\rm eff}$ is twice the best fit value. The wind velocity is fixed at 50 $\kms$.
Solid curves: globular clusters. Long-dashed curves: dwarf spheroidals. Short-dashed curves: nuclear star clusters. Dotted curves: early-type galaxies. Bottom panel: accretion radius for the same systems.  
}
\label{size}
\end{figure}

In Figure~\ref{size} we instead fix $v_{\rm wind}$, and allow for a
scatter in the mass-size relationship. For globular clusters we assume
$R_{\rm eff}=1$ pc; $R_{\rm eff}=2$ pc and $R_{\rm eff}=4$ pc.  For
galaxies, the middle curve shows the best fit $R_{\rm eff}$ for a given stellar
mass value (Equations 1, 2 and 3), the top curves assume that $R_{\rm
  eff}$ is half the best fit value, and the bottom curves assume that $R_{\rm
  eff}$ is twice the best fit value. We have assumed $M_{\rm stellar}=
10^5-10^7 M_\odot$ for globular clusters, $M_{\rm stellar}= 10^5-10^8
M_\odot$ for dwarf spheroidals and nuclear star clusters, and $M_{\rm
  stellar}= 10^8-10^{11} M_\odot$ for early type galaxies, limiting
our investigation to the mass ranges probed by \cite{Shen2003,Walker2009,Seth2008}.  In this plot the $v_{\rm
    wind} \gg \sigma$ limit of Equation~6 becomes evident: at low stellar masses, for every
  type of stellar distribution but for the early type galaxies,
  $R_{\rm acc}$ does not depend on $R_{\rm eff}$, and it is determined
  only by the BH mass and the assumed $v_{\rm wind}$. The early type
  galaxies generate deeper potential wells, never reaching the $v_{\rm
    wind} \gg \sigma$ limit.

The bolometric luminosity of the massive black hole can be written as:
\beq
L_{\rm bol}=\epsilon \dot{M} c^2\;,
\label{eq:lbol}
\eeq where $\epsilon$ represents the fraction of the accreted mass
that is radiated away. The nature of the accretion process, and the consequent value of $\epsilon$, is rather uncertain.  AGNs accrete through accretion discs with a high efficiency ($\epsilon\sim 0.1$). Supermassive BHs at the centers of quiescent galaxies, including the Milky Way, can have luminosities as
low as $\sim 10^{-9}-10^{-8}$ of their Eddington values (e.g. \cite{Loewenstein2001}), and well below the luminosity one would estimate assuming $\epsilon\sim 0.1$.  

Following \cite{Merloni2008} we define  $\lambda \equiv  L_{\rm bol}/L_{\rm  Edd}$, where $L_{\rm bol}$ is the bolometric luminosity and $L_{\rm  Edd}=4 \pi G M_{\rm BH} m_{\rm p} c / \sigma_{\rm T} \simeq 1.3 \times 10^{38} (M_{\rm  BH}/M_{\odot})$ erg s$^{-1}$ is the Eddington luminosity. We write the radiative efficiency, $\epsilon$, as a combination of the accretion efficiency,
$\eta$, that depends only on the location of the innermost stable circular orbit\footnote{If the
  viscous torque vanishes at the innermost stable circular orbit, then   $\eta$ is a function of BH spin only, ranging from $\eta\simeq
  0.057$ for Schwarzschild (non-spinning) black holes to $\eta\simeq   0.42$ for maximally rotating Kerr black holes.}, here assumed to be $\eta=0.1$, and of a term, $\eta_{\rm acc}$,  that
depends on the properties of the accretion flow itself: $\epsilon=\eta\,\eta_{\rm acc}$. We also define $\dot m=\eta \dot M c^2/L_{\rm Edd}$.

\begin{figure}
\plotone{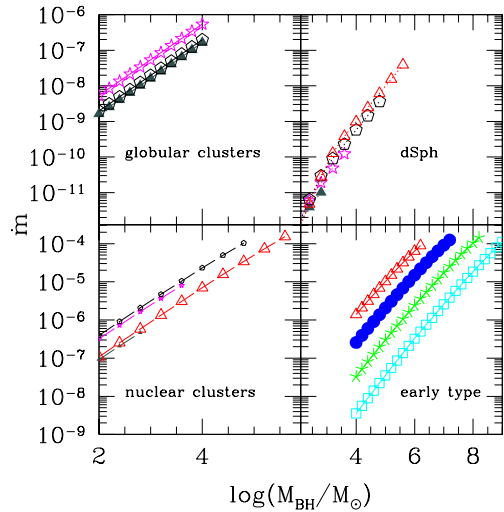}
\caption{\footnotesize Accretion rate, in Eddington units, of massive black holes in different stellar systems. 
	Top right: dwarf spheroidals;  bottom right=early type galaxies; bottom left=nuclear clusters; top left=globular clusters.
	Gray filled triangles: $M_{\rm stellar}=10^{5} M_\odot$; magenta stars=$M_{\rm stellar}=10^{6} M_\odot$; 
	black pentagons =$M_{\rm stellar}=10^7 M_\odot$; red empty triangles=$M_{\rm stellar}=10^8 M_\odot$; blue dots=$M_{\rm stellar}=10^9 M_\odot$;
	green asterisks=$M_{\rm stellar}=10^{10} M_\odot$; cyan squares=$M_{\rm stellar}=10^{11} M_\odot$.
	The mass-size relationship is given by Equations 4, 5 and 6. We assume $R_{\rm eff}=2$ for globular clusters. The wind velocity is $v_{\rm wind}=50 \kms$.}
\medskip
\label{fedd}
\end{figure}

For `radiatively efficient' accretion, $\eta_{\rm acc}=1$. To estimate the X--ray luminosity, we apply a simple bolometric correction, and assume that the X-ray luminosity is a fraction $\eta_{\rm X}$ of the bolometric luminosity.  Ho et al. (1999) suggest that for low-luminosity AGN, with Eddington rates between $10^{-6}$ and $10^{-3}$ the luminosity on the [0.5-10] keV band represents a fraction 0.06-0.33 of the bolometric luminosity.  We assume here $\eta_{\rm X}=0.1$, so that $L_{\rm X}=\eta_{\rm X}\, \epsilon\, \dot M c^2$, where $\epsilon=\eta=0.1$. We refer to this model as `radiatively efficient'.  

Since the accretion rates we find are very sub-Eddington, we assume, in a second model, that the accretion flow is optically thin and geometrically thick.  In this state the radiative power is strongly suppressed \citep[e.g.,][]{Narayan1994,Abramowicz1988}.   \cite{Merloni2008} suggest that this transition occurs at $\dot m<\dot m_{\rm cr}=3\times 10^{-2}$, and that  $\eta_{\rm acc}=(\dot m/\dot m_{\rm cr})$, so that $\epsilon=\eta(\dot m/\dot m_{\rm cr})$. The X-ray luminosity is therefore: $L_{\rm   X}=\eta_{\rm X}\, \epsilon\, \dot M c^2$, where again $\eta_{\rm X}=0.1$. We refer to this model as ``radiatively inefficient".

\begin{figure}
\plotone{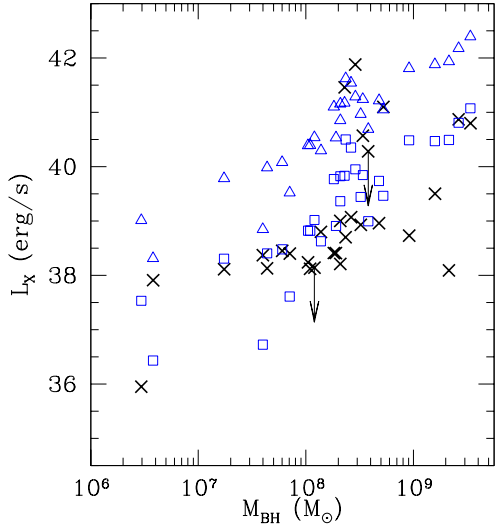}
\caption{\footnotesize X--ray luminosity (top) of massive black holes in 29 nearby elliptical galaxies. Crosses and upper limits are from from Pellegrini (2010), where we select only black holes with dynamical mass measurement. Triangles: ``radiatively efficient" model. Squares: ``radiatively inefficient" model. }
\medskip
\label{LE}
\end{figure}

In Figure~\ref{fedd} we show the accretion rate, in Eddington units, when we assume $\eta=0.1$. Hereafter we vary the mass of the massive black hole from $100\msun$ to $10^4\msun$ for globular clusters, since there is no firm conclusion that massive black holes' masses scale with the mass of the stellar component as $M_{\rm BH}=10^{-3} M_{\rm stellar}$. For galaxies we assume instead an upper limit to the massive black hole mass corresponding to  $M_{\rm BH}=2 \times 10^{-2} M_{\rm stellar}$, a lower limit of $100\msun$ for dSph and nuclear clusters and a lower limit of $10^4\msun$ for early type galaxies. 

 We complete the exercise by adding observational results for a sample of 29 early type galaxies where both dynamical black hole mass and  X--ray luminosity \citep{Pellegrini2010}\footnote{The sample of Pellegrini (2010) comprises 112 galaxies with measured X-ray luminosity. For systems that do not have dynamical black hole mass measurement, Pellegrini (2010) derives black hole masses from the M-$\sigma$ relation.  We limit our analysis to those galaxies that have a direct black hole mass measurement to avoid adding additional uncertainties, especially below $M_{\rm BH}=10^7 \msun$, where the   M-$\sigma$ relation is less secure. We note, however, that the results we discuss hold for the whole sample.} are available \citep[see also][]{Soria2006a,Gultekin2009b}. 24 of these galaxies also report the stellar mass of the bulge \citep{MarconiHunt2003}. For those galaxies where the bulge mass is unavailable we derive stellar masses from B-band magnitudes. For these galaxies we also derive B-band luminosities directly from $L_V$ \citep{Gultekin2009}, assuming $B-V=1$ \citep{Coleman1980}, and we check that our choice of a mass-to-light ratio of 5 agrees well with this complementary technique to derive $L_B$. 

Figure~\ref{LE} compares the luminosities we predict for these galaxies to the measured X-ray luminosity of the galaxies (or upper limits). In agreement with the conclusions of \cite{pellegrini2005} and \cite{Soria2006b} the radiatively inefficient case best fits the luminosity of most systems, except the most luminous ones. Overall, even the radiatively inefficient case slightly overestimates the luminosity, at least at the high mass end, and we find that, for instance,  $\eta_{\rm X}=0.03$ provides a much better fit.  As discussed by Pellegrini (2010) there seems to be a smooth transition between radiatively inefficient and radiatively efficient accretion.

We also estimate the X--ray luminosities for Milky Way dSphs with stellar mass $>10^5 \, \msun$, where we use directly $R_{\rm half}$ and $M_{\rm stellar}$ from \cite{Walker2010}. We assume in one case that $M_{\rm BH}=10^{-3} M_{\rm stellar}$, and in another case that black holes have a fixed massive black hole mass of $10^5 \msun$, based on models presented in \cite{svanwas2010}.  
We note that in all these cases the X--ray luminosities for massive black holes in dwarf galaxies are below $10^{35} {\rm erg\, s^{-1}}$.  Figure~\ref{lum} summarizes our primary results; predicted X-ray luminosities for different stellar systems. 

\section{Discussion}

We have developed a simple model to estimate the level of accretion fueled by recycled stellar winds on black holes hosted in stellar systems of different types.  Let us examine the various assumptions of our models to question if our approach is too conservative.  To model the accretion rate we need (1) a stellar density profile, (2) physical size and mass of a system, (3) a total mass loss from stars (which depends on their age and luminosity), and (4) a velocity of stellar wind. 

Regarding points (3) and (4), we note that our choice of stellar ages and mass-to-light ratios are already quite optimistic (except for the case of globular clusters and early type galaxies, but we note that our results for globulars are consistent with the estimate of Miller \& Hamilton 2002), and for most massive black holes in massive stellar systems the wind velocity is not highly influential. Regarding point (2), we can see from Fig.~\ref{size} that the relationship between size and radius does not have a very strong effect on our results. More interesting is point (1). As long as the wind velocity is larger than the velocity dispersion of a galaxy, the amount of available gas will increase if the density profile is steeper. For instance, an ideal density profile is an isothermal sphere (possibly singular) where the velocity dispersion is constant while the central density increases towards the center. In such case the size of the accretion radius, and the accretion rate, are maximized (see Equations~6 and 7).

\begin{figure*}
\plottwo{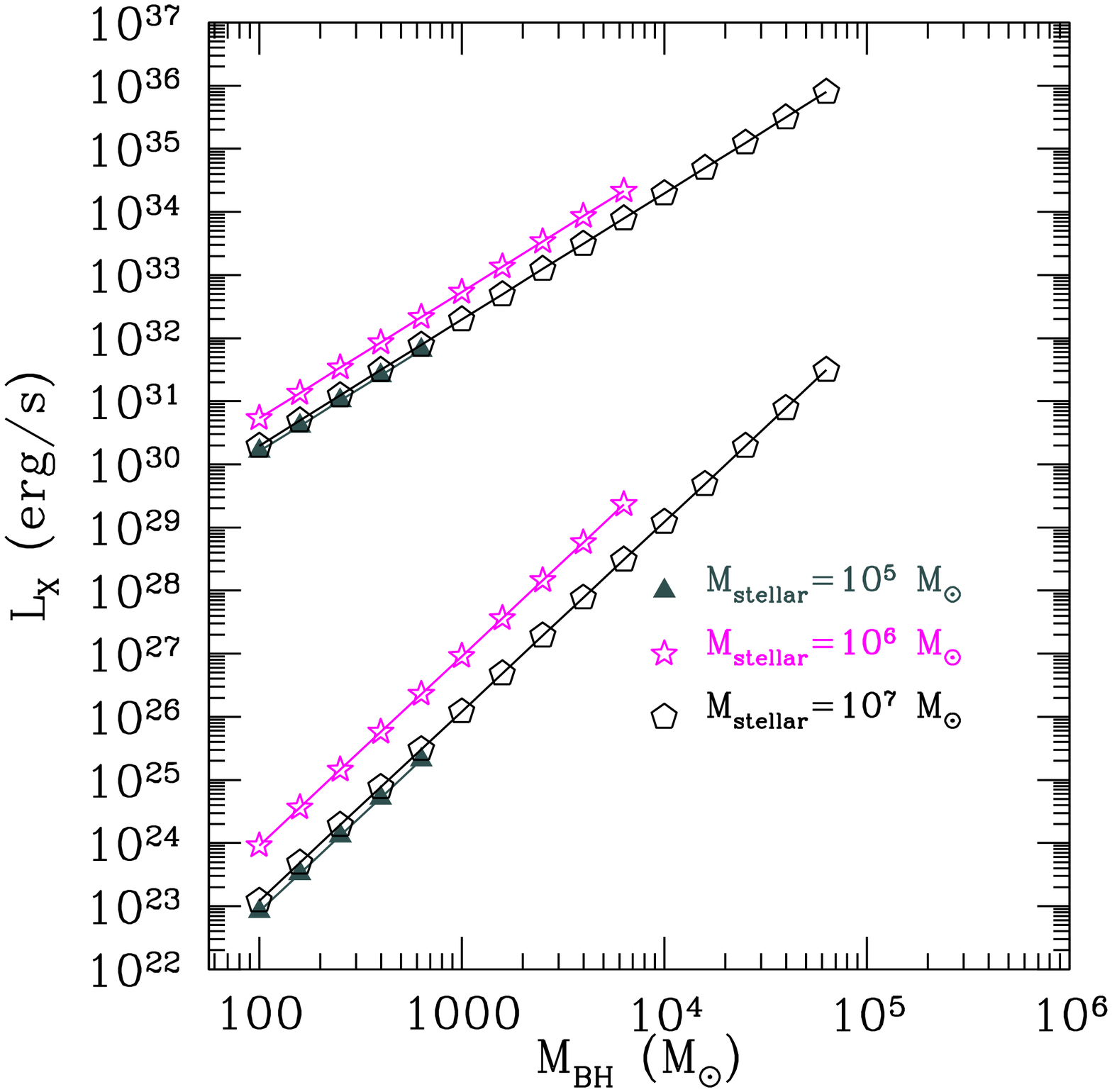}{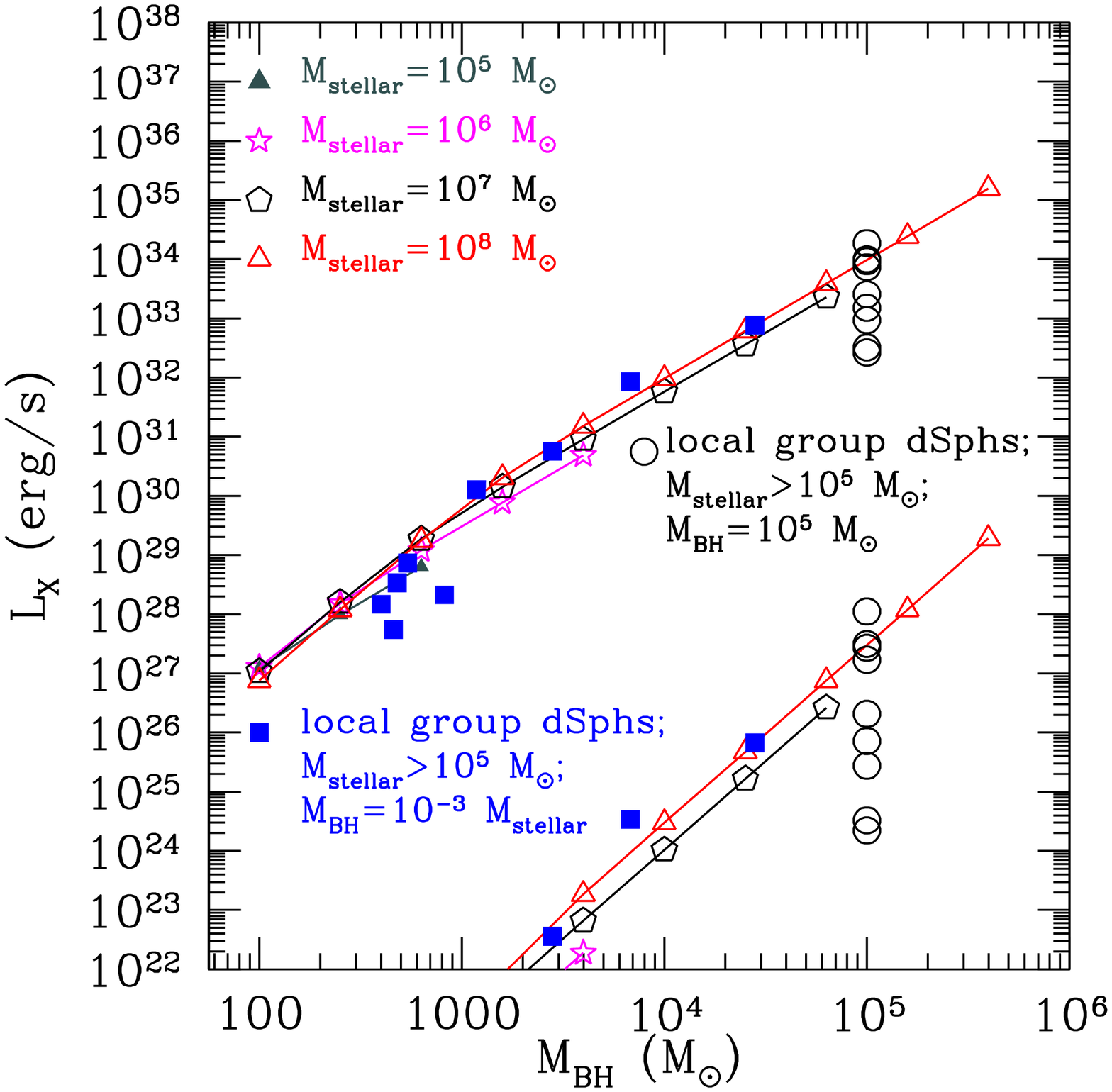}
\plottwo{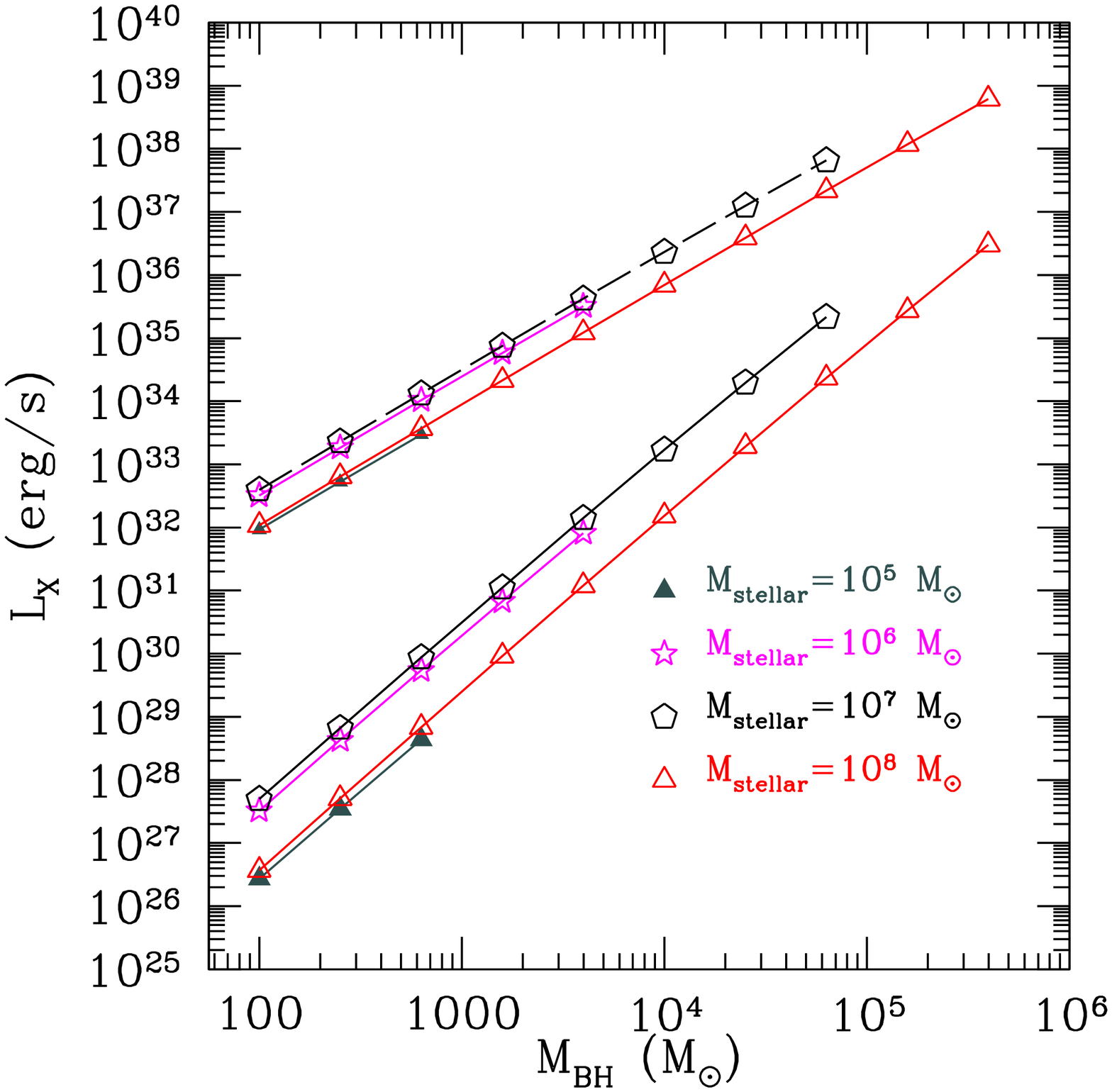}{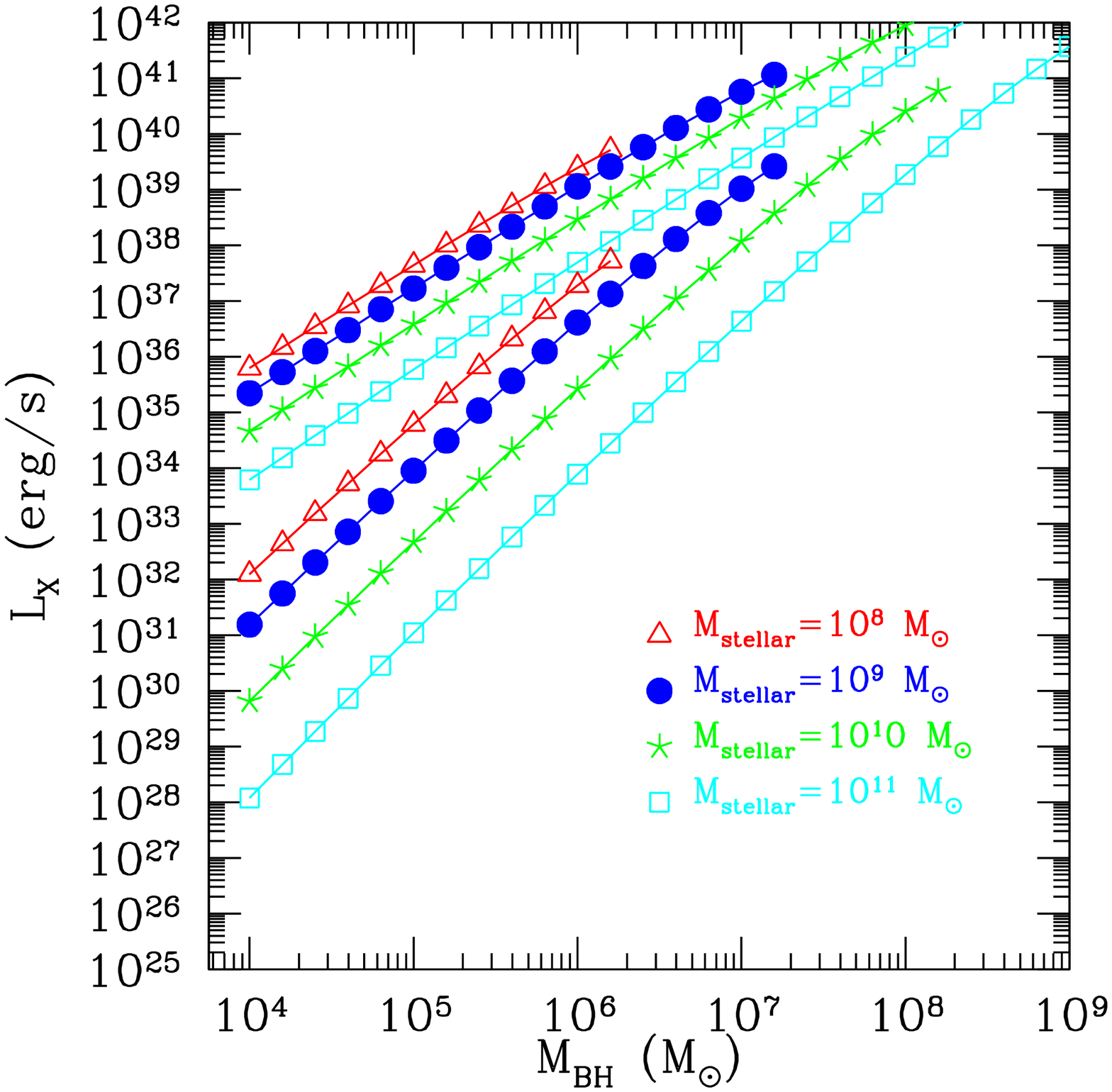}
\caption{X-ray luminosity of massive black holes in different stellar systems, assuming a radiatively efficient accretion flow (top series of curves) and a radiatively inefficient accretion flow (bottom series of curves). The wind velocity is $v_{\rm wind}=50 \kms$. Top-left: globular clusters. Top-right: dwarf spheroidals. (Squares: luminosity we derive for dSphs with stellar mass $>10^5 \, \msun$ assuming massive black holes with mass $10^{-3}$ times the stellar mass, and using dynamical masses and radii from Walker et al. 2009.  Open circles: luminosity a $M_{\rm BH}=10^5 \msun$ massive black hole would have in the same galaxies). Bottom-left: nuclear clusters. Bottom-right: early-type galaxies.
	Gray filled triangles: $M_{\rm stellar}=10^{5} M_\odot$; magenta stars: $M_{\rm stellar}=10^{6} M_\odot$; 
	black pentagons: $M_{\rm stellar}=10^7 M_\odot$; red empty triangles: $M_{\rm stellar}=10^8 M_\odot$; blue dots: $M_{\rm stellar}=10^9 M_\odot$;
	green asterisks: $M_{\rm stellar}=10^{10} M_\odot$; cyan squares: $M_{\rm stellar}=10^{11} M_\odot$. We have assumed $M_{\rm stellar}= 10^5-10^7 M_\odot$ for globular clusters; $M_{\rm stellar}= 10^5-10^8 M_\odot$ for dwarf spheroidals and nuclear star clusters, and $M_{\rm stellar}= 10^8-10^{11} M_\odot$ for early type galaxies, limiting our investigation to the mass ranges probed by Shen et al., Walker et al. and Seth et al.}
\medskip
\label{lum}
\end{figure*}

One of our goals was to assess the detectability of putative massive black holes in Milky Way dSphs.  If they exist they provide valuable information on the process that formed the seeds of the massive holes we detect in much larger galaxies \citep{svanwas2010}.  Figure~\ref{lum} suggests that such black holes would be elusive, as the expected luminosities are often even less than those of X-ray binaries.  Regarding the three points discussed above,  in the case of dSphs,  the observed stellar density profiles are very shallow  and the central stellar densities low, of order of at most a few stars per cubic parsec \citep{Irwin1995}. We therefore consider the choice of a steeper profile inappropriate. The analytical fit of the mass-size relationship (eq. 4) could be inaccurate, but as we show in Fig.~\ref{lum}, where we model specific galaxies using their measured masses and radii, we find that the luminosities are in very good agreement with what we find using the analytical fit.  Finally, even assuming a wind velocity $v_{\rm wind}=10 \kms$ the luminosities are always below $10^{38}$ erg/s, even pushing the stellar age to 1 Gyr. 

On the other hand, our model could indeed be too conservative for the case of nuclear star clusters. These systems have a wide spread in stellar ages \citep[e.g.,][]{Carollo2001} and they exhibit steep density profiles (see \cite{Kormendy2004} for a comprehensive review). Decreasing the typical stellar age to 1 Gyr increases the luminosity by about a factor 10, while modeling the density profile as a singular isothermal sphere increases the luminosity by several orders of magnitude, with black holes with masses $>10^4 \msun$ hosted in nuclear clusters with mass $>10^7 M_\odot$ attaining luminosities $>10^{38}$ erg/s. At luminosities below $>10^{38}$ erg/s the contamination from X-ray binaries is high (Gallo et al. 2010 and references therein) and we consider this to be a `safe' threshold for black hole detection.  Surveys built in the same spirit of AMUSE-Virgo and AMUSE-field, with a limiting luminosity of order $>10^{38}$ erg/s are likely to provide an excellent venue to test our models. 

\acknowledgements 
We would like to thank E. Gallo and D. Maitra for fruitful discussions, and M. Walker for clarifications on his data.

\end{document}